\documentstyle[aps]{revtex}
\begin{document}
\baselineskip 1cm
\title{Soliton Squeezing in a Mach-Zehnder Fiber Interferometer}

\author{Marco Fiorentino, Jay E. Sharping, and Prem Kumar}
\address{Center for Photonic Communication and Computing,
Department of Electrical and Computer Engineering, Northwestern University, 
2145 N. Sheridan Road, Evanston, IL 60208-3118, USA}

\author{Dmitry Levandovsky}
\address{Tellabs Operations Inc., 4951 Indiana Ave., Lisle, IL 60532, USA}

\author{Michael Vasilyev}
\address{Corning Inc., 2200 Cottontail Lane, Sommerset, NJ 08873, USA}
\maketitle
\begin{abstract}

A new scheme for generating amplitude squeezed light by means 
of soliton self-phase modulation is experimentally 
demonstrated. By injecting 180-fs pulses into an equivalent 
Mach-Zehnder fiber interferometer, a maximum noise 
reduction of $4.4 \pm 0.3$ dB is obtained ($6.3 \pm 0.6$ dB when 
corrected for losses). 
The dependence of noise reduction on the interferometer splitting ratio 
and fiber length is studied in detail.  

\end{abstract}


\noindent

Current interest in efficient generation of squeezed radiation in
optical-fiber devices is prompted by the possibility of using the
resulting highly-squeezed fields to create and distribute
continuous-variable entanglement for quantum-communication applications
such as teleportation and dense coding~\cite{telep}. Several schemes to
generate amplitude-squeezed light by use of the Kerr nonlinearity in
optical fibers have recently been devised
\cite{Rosenbluh,Schmitt,Spalter,Krylov,Dm}. The most successful soliton
amplitude-squeezing experiments to date have used an asymmetric Sagnac
interferometer, leading to 3.9 dB (6.0 dB when corrected for losses)
\cite{Schmitt} and 5.7 dB (6.2 dB when corrected for losses)
\cite{Krylov} of noise reduction, respectively. As shown schematically
in Fig.~\ref{ball}, the amplitude squeezing in such experiments results
from the combined effect of soliton's nonlinear propagation in the fiber
and its subsequent mixing with an auxiliary weak pulse at the fiber
output. The mixing rotates the total mean field relative to the
soliton's noise, which gets shaped into a ``crescent'' through the
nonlinear interaction, thus allowing the squeezing to be measured in
direct detection. To obtain a high degree of noise reduction, it is then
crucial to control the relative amplitude and phase of the two pulses.
In the Sagnac configuration~\cite{Schmitt,Krylov} the pulses counter
propagate in the same fiber, thus preventing the control of their
relative phase. Moreover, the relative amplitude between the
soliton-like pulse and the auxiliary dispersive pulse is unchangeable
once one fixes the splitting ratio of the interferometer beamsplitter.

To overcome these drawbacks we investigate the setup shown in
Fig.~\ref{schematic}. In this rendition, the Sagnac geometry is
``unfolded'' into a polarization Mach-Zehnder formed between two
polarization beam splitters, PBS2 and PBS3, wherein the two arms of the
interferometer correspond to the two polarization modes of a single-mode
polarization-maintaining (PM) fiber (3M-FSPM 7811). The Mach-Zehnder
geometry allows independent control of the phase and amplitude of the
auxiliary pulse relative to the soliton. By adding these new degrees of
freedom, we are able to study in greater detail the physical process
that leads to noise reduction. 

In our setup the light source is a tunable optical-parametric oscillator
(Coherent Inc., model Mira-OPO) emitting a train of pulses at a
wavelength of 1538\,nm with a 75MHz repetition rate and 180fs (FWHM)
pulse width. The pulses are $sech$ shaped and nearly Fourier transform
limited (time-bandwidth product $\simeq $ 0.4). The source is used to
inject one arm of the interferometer with a strong pulse, propagating in
the soliton regime, and the other with a weak, dispersive pulse ($\simeq
$ 10\% of the fundamental soliton energy). The total injected power and
the relative powers of the two pulses are controlled by using a
half-wave plate (HWP1) and a polarizing beamsplitter (PBS1). Since the
pulses propagate with significantly different group velocities in the
two polarization modes, they are launched at different times into the
fiber in order for them to overlap at the fiber output. The relative
delay is introduced by adding separate free-space propagation paths
[$s(p)$-polarization reflects from M1 (M2)] for the two polarization
modes in the interferometer. This arrangement also prevents cross
interaction between the two pulses, as they are temporally separated
during most of the propagation distance in the fiber. In addition, a
piezoelectric control on M1 allows fine tuning of the relative phase
between the pulses. A half-wave plate (HWP2) and a quarter-wave plate
(QWP3) are used to inject the $s$ and $p$ polarized pulses from free
space into the correct polarization modes of the fiber.\footnote{In
principle QWP3 shouldn't be necessary, but we found that it greatly
improves the polarization purity of the pulses at the fiber output.
This, perhaps, is due to the fact that the polarization eigenmodes of
the PM fiber are not exactly linearly polarized.} At the output of the
fiber, the two pulses are recombined using a half-wave plate (HWP3) and
a polarizing beamsplitter (PBS3), allowing us to easily change the
recombination ratio $T$ (defined as the ratio of powers of the weak to
the strong pulse reflected by PBS3) by turning HWP3. The combined pulse,
reflected by PBS3, is reflected off another polarization beamsplitter
(PBS4) in order to insure a high polarization purity while minimizing
optical losses. The emerging 75MHz pulse train is then analyzed with a
four-diode balanced detector, wherein four photodiodes (Epitaxx ETX500)
are used to increase the saturation power to 40\,mW of average power. To
increase the overall detection efficiency, we use spherical mirrors that
bounce back the light reflected from the photodiode surfaces. This
configuration yields a total measured detection efficiency of 78\%,
where the Fresnel loss at the fiber end (5\%), propagation losses
through various optical elements (8\%), and sub-unity photodetector
quantum efficiencies (equivalent losses of 11\%) are the contributing
factors. For better squeezing measurements, the Fresnel loss was also
mitigated by using a window that is anti-reflection coated on one side.
The uncoated side is put in optical contact with the uncoated fiber tip
using an index-matching gel (Nye Lubricants OC-431A-LVP). We estimate
that this improves the overall detection efficiency to 82\%.

The experimental results are shown in Fig.~\ref{exp}. We study the
quantum-noise reduction in the photocurrent spectral density at
28.7\,MHz as a function of the average optical power which is
proportional to the squared soliton number $N^2$ in the soliton arm of
the interferometer. Figures~\ref{exp}(a)--(c) illustrate the
quantum-noise reduction for fiber lengths of 3.4, 6, and 9\,m,
corresponding to 2.5, 4.3, and 6.5 soliton periods, respectively. The
noise reduction was maximized by adjusting the relative phase between
the pulses in the interferometer arms. Data sets represented by filled
circles refer to the maximum noise reduction, which is obtained with $T$
equal to: (a) $(0.4\pm 0.1)\% $, (b) $(0.32\pm 0.01)\% $, and (c)
$(0.53\pm 0.05)\% $, respectively. The best noise reductions measured
are: $(4.1\pm 0.3)$\,dB for the setup with 78\% detection efficiency and
$(4.4\pm 0.3)$\,dB for that with 82\% detection efficiency. These
correspond respectively to $(6.3\pm 0.6)$\,dB and $(6.6\pm 0.7)$\,dB of
inferred noise reduction at the fiber output when degrations due to
linear losses are taken into account.

The data in Fig.~\ref{exp} clearly show a pronounced dependence of the
noise reduction on the recombination ratio $T$, the fiber length, and
the input average power. The analysis of such dependencies requires a
model for the propagation of the pulses' quantum noise in the fiber. An
analytical solution for the propagation of the quantum noise of the
fundamental soliton ($N=1$) and the associated continuum is
possible~\cite{Haus,Kaup,Dmitry}. This analytical solution has been
applied to the analysis of squeezing experiments that use
interferometers~\cite{Dmitry}. We show the comparison of this latter
analysis with our experimental data in the inset of Fig.~\ref{num}(a),
where only the points with $N=1$ are plotted. However, when the pulses
are not fundamental solitons, there is no known analytical solution for
the propagation of the quantum noise. Several works have applied
numerical methods to solve this problem and such numerical solutions
have been used to analyze the noise-reduction properties of asymmetric
Sagnac interferometers \cite{Schmitt,Doerr}. A common feature of all of
these works is that the calculated noise reduction exceeds, by a sizable
amount, any experimental result obtained so far. Remarkably, however,
the theory seems to be able to reproduce the overall structure of the
experimental data.

In our analysis, we have focused on the features of quantum-noise
reduction vs.\ average power and the associated dependence on the
recombination ratio. To do this, we have performed a numerical
simulation of the Mach-Zehnder interferometer. At the core of the
simulation is a routine that propagates quantum noise through the fiber.
This routine relies on the standard linearization technique
\cite{Shirasaki} for solving the quantum nonlinear Schr\"{o}dinger
equation (NLSE)
\begin{equation}
\frac{\partial \hat U}{\partial z}=i \hat U^{\dagger} \hat U \hat U + 
\frac{i}{2} \frac{\partial^2 \hat U}{\partial t^2},
\label{NLSE}
\end{equation}     
which describes the evolution of the annihilation operators $\hat U$ for
the envelope of the electric field in the fiber. Equation~(\ref{NLSE})
is written in a retarded frame moving together with the pulse along $z$,
which is expressed in standard normalized-length units~\cite{Agrawal}.
We linearize this equation by putting $\hat U=\langle{\hat U}\rangle +
\hat u$, where $\hat u$ is the annihilation operator for the
fluctuations, keeping only terms up to first order in $\hat{u}$. This
yields a pair of coupled equations: the zeroth-order expansion
represents the classical NLSE, which describes the evolution of the
envelope $\langle{\hat U}\rangle$, whereas the first-order expansion
gives a linear equation for the fluctuation operator $\hat{u}$. Applying
the discretization procedure described in Ref.~\cite{Doerr}, we have
developed a computer program to solve the fluctuation equation
numerically, assuming that a solution (numerical or analytical) is given
for the corresponding classical NLSE. Our algorithm uses a split-step
Fourier method, wherein each step is solved exactly using matrix
exponentiation. The behavior of the Mach-Zehnder interferometer is
simulated by propagating the averaged envelopes of the soliton-like and
the auxiliary pulses through equal lengths of the fiber. Using these
numerical solutions for the averaged envelopes we then propagate the
noise operators for the soliton-like pulse. At the output of the
interferometer, the soliton-like pulse is mixed with the auxiliary
pulse. We consider two cases in our simulations: for the curves marked
by hollow squares connected with a dashed line, we assume coherent-state
noise for the auxiliary pulse; whereas for the curves marked by solid
circles connected with a continuous line, we numerically propagate the
noise of the auxiliary pulse through the fiber as well. As expected, the
effect of the quantum-noise evolution of the auxiliary pulse is larger
for larger recombination ratios [Fig. \ref{num} (c)]. We attribute the
oscillations in the noise reduction to phase chirp introduced by
dispersion in the fiber. This also explains why the frequency of
oscillations is approximately doubled, when the propagation of the
auxiliary pulse in the fiber is fully taken into account.

In Fig.~\ref{num}(a)--(c) we compare the results of our simulations for
the 6.0\,m fiber length and various recombination ratios with the
corresponding experimental data sets that have been corrected for
detection losses. The calculated values of noise reduction do not
quantitatively agree with the experimental data, which show little
evidence of the oscillations. Nevertheless, the theory correctly
predicts a saturation in the detected noise reduction for large values
of $N^2$, which we attribute to increasing temporal mismatch in the
overlap between the soliton-like and the auxiliary pulses. Reasons for
the discrepancy between the theory and the experiment have been
discussed in the literature with authors reaching differing conclusions.
Some~\cite{Schmitt} believe the discrepancy to be due to the Raman
effect (that we have found to be significant in our setup) and their
claim is partially supported by papers that discuss limits on squeezing
imposed by the Raman noise~\cite{Shapiro}. Other authors~\cite{Werner}
have reported results of numerical calculations with inclusion of the
Raman effect, but the results show that the Raman noise does not seem to
play a significant role. Finally, the authors of one
experiment~\cite{Rosenbluh} cooled the fiber to liquid nitrogen
temperature and observed an increase in the noise reduction, thus
suggesting a preeminent role for thermal phase scattering owing to
guided-wave Brillouin scattering; whereas others~\cite{Bergman} were
able to achieve a good matching between the theory and data in an
experiment where both Brillouin and Raman scattering were negligible.
These discrepancies between the various studies suggest that further
investigation, both theoretical and experimental, is warranted.

In conclusion, we have studied a Mach-Zehnder nonlinear fiber
interferometer for generation of amplitude-squeezed light. By varying
the key parameters of the interferometer, such as the fiber length,
recombination ratio, and average input power, we optimized its
performance and measured large amounts of amplitude squeezing. In
addition, by using numerical analysis we have tried to understand the
dependencies of noise reduction on these key parameters.

The authors thank G. Biondini for useful suggestions in the numerical
simulations and E. Corndorf for his help with the computers. This work
was supported in part by the U.S. Army Research Office through the MURI
grant DAAD19-00-1-0177 and the associated MURI Fellowship
(DAAD19--00--1--0469) for J. E. Sharping.

\newpage
\section*{List of figures}

\begin{figure}[h]
\caption{A schematic of the mechanism leading to amplitude squeezing in
nonlinear fiber interferometers. (a) Amplitude dependent phase shift
deforms the solitons' symmetrical coherent-noise distribution into a
crescent-shaped distribution. (b) The auxiliary pulse is used to rotate
the mean field relative to the crescent to allow maximum amplitude-noise
reduction to be measured in direct detection.}
\label{ball}
\end{figure}

\begin{figure}[h] 
\caption{A schematic of the experimental setup. The shaded area
highlights the components that form the Mach-Zehnder interferometer.
HWP, half-wave plate; QWP, quarter-wave plate; PBS, polarizing
beamsplitter; M, mirror. The inset shows a typical plot of the
sum-photocurrent noise as the auxiliary-field phase is scanned; the
bottom plot is obtained by fixing the phase to maximize the noise
reduction. Both plots were obtained by averaging 5 time scans.}
\label{schematic}
\end{figure}

\begin{figure}[h] 
\caption{Experimental quantum-noise reduction as a function of average 
power in the soliton arm (squared soliton number $N^2$). 
(a) 3.4 m of fiber, $T = (0.4\pm 0.1)\%$; 
(b) 6.0 m of fiber, 
$T = (0.11\pm 0.01)\%$ (triangles),
$T = (0.32\pm 0.01)\%$ (filled circles), 
$T = (1.0\pm 0.4)\%$ (squares), and 
for clarity only one error bar for each data set is displayed;
(c) 9 m of fiber, $T = (0.53\pm .05)\%$. 
For all data the detection efficiency is 78\%.}
\label{exp}
\end{figure}

\begin{figure}[h] 
\caption{Comparison of the experimental data, corrected quantum-noise 
reduction vs.\ the squared soliton number $N^2$ (diamonds), with 
numerical simulations: hollow squares refer to the case where 
the auxiliary pulse is in a coherent state and filled circles to 
the case where input quantum noise of the auxiliary pulse is 
propagated through the fiber; lines connecting the points are 
added for guidance.  
(a) 6.0 m of fiber, 
$T = (0.11\pm 0.01)\%$ (experiment), $T = 0.1\%$ (simulations);
(b) 6.0 m
of fiber, $T =(0.32 \pm 0.1)\%$ (experiment) $T = 0.35 \%$ (simulations);
(c) 6.0 m of fiber, 
$T = (1.0\pm 0.4)\%$ (experiment), $T = 1.0\%$ (simulations). 
In the inset in (a) shows a comparison of the theoretical results of
Ref.~[10] (down triangles) with the corrected experimental quantum-noise
reductions (up triangles) in cases where the strong pulse is a
fundamental soliton. The data points are for various fiber lengths with
correspondingly different recombination ratios $T$.}
\label{num}
\end{figure}
\end{document}